\documentclass[%
 reprint,
superscriptaddress,
nofootinbib,
 amsmath,amssymb,
 aps,
prd,
floatfix,
]{revtex4-2}

\usepackage[colorlinks=true
,urlcolor=blue
,anchorcolor=blue
,citecolor=blue
,filecolor=blue
,linkcolor=blue
,menucolor=blue
,linktocpage=true
,pdfproducer=medialab
,pdfa=true]{hyperref}
\usepackage{cleveref} 

\crefname{equation}{Eq.}{Eqs.}
\crefname{section}{Section}{Sections}
\crefname{figure}{Fig.}{Figs.}
\crefname{table}{Table}{Tables}
\crefname{appendix}{Appendix}{Appendices}

\usepackage[caption=false]{subfig}
\usepackage{aas_macros}
\usepackage{amsmath}
\usepackage{caption}
\usepackage{ragged2e}
\usepackage{adjustbox}
\usepackage{xcolor}
\usepackage{siunitx}
\usepackage{graphicx}
\usepackage{dcolumn}
\usepackage{bm}

\begin{document}

\preprint{APS/123-QED} 
\title{Finding Strongly Lensed Supernovae from Blended Light Curves}

\author{Sangwoo Park}
\email{swpark@kasi.re.kr}
\author{Arman Shafieloo}
\email{shafieloo@kasi.re.kr}
\affiliation{Korea Astronomy and Space Science Institute (KASI), 776 Daedeok-daero, Yuseong-gu, Daejeon 34055, Republic of Korea}
\affiliation{University of Science and Technology, 217 Gajeong-ro, Yuseong-gu, Daejeon 34113, Republic of Korea}

\author{Alex G. Kim}
\affiliation{Lawrence Berkeley National Laboratory, 1 Cyclotron Road, Berkeley, CA 94720, USA}

\author{Eric V. Linder}
\affiliation{Berkeley Center for Cosmological Physics \& Berkeley Lab, University of California, Berkeley, CA 94720, USA}

\author{Xiaosheng Huang}
\affiliation{Department of Physics \& Astronomy, University of San Francisco, San Francisco, CA 94117-1080, USA}

\date{\today}

\begin{abstract}

We present a model-independent, photometry-only framework for identifying strongly lensed supernovae when multiple images are unresolved and blended into a single point source. Building on the simulation-based methodology of~\cite{Bag:2020pbg}, we apply this approach to real Zwicky Transient Facility (ZTF) data using a validation sample of spectroscopically confirmed Type~Ia supernovae. The method models the observed flux as a superposition of two time-shifted components, and Bayesian inference is used to estimate the relative scaling and time delay.
Applying this framework to 445 well-converged supernovae, we find that only a single object satisfies the selection criteria when adopting a conservative threshold of $\Delta t \ge 12$ days, corresponding to a false positive fraction of $1/445 \approx 0.22\%$. 
A laxer threshold of $\Delta t \ge 10$ days yields fourteen objects, for a false positive fraction of $3.15\%$. 
The method provides a scalable and model-independent first-stage filter for identifying lens-like candidates in large time-domain surveys such as the Rubin Observatory's Legacy Survey of Space and Time (LSST).

\end{abstract}

\maketitle

\section{Introduction}
\label{introduction}

Strong gravitational lensing of supernovae (SNe) provides a unique opportunity to probe both cosmology and astrophysics~\cite{Oguri:2010ns}. When a background supernova is lensed by a massive foreground object such as a galaxy, multiple images of the supernova can form, separated by time delays of days to weeks. These time delays encode information about the cosmic distance scale and the Hubble constant $H_0$, as first proposed by Refsdal~\cite{1964MNRAS.128..307R}, and further developed into a practical tool for time delay cosmography~\cite{Linder:2004hx,Linder:2011dr,Suyu:2012aa,Treu:2016ljm,H0LiCOW:2019pvv}. Strongly lensed supernovae also enable lens mass modeling~\cite{Grillo:2018ume} and offer rare windows into high-redshift supernovae that would otherwise be too faint to observe~\cite{Kelly:2014mwa,Goobar:2016uuf,Petrushevska:2017kza}.

Despite their scientific potential, confirmed strongly lensed supernovae remain extremely rare. To date, only a few have been securely identified, such as SN Refsdal~\cite{Kelly:2014mwa}, iPTF16geu~\cite{Goobar:2016uuf}, and SN H0pe~\cite{Pascale:2024qjr}, often discovered via high-resolution imaging, redshift inconsistencies, or careful lens modeling~\cite{XXX:2014xxi,Rodney:2015uyq,Rodney:2021keu,Goobar:2022wan}. These methods, while effective, are observationally expensive and not easily scalable to large photometric surveys.

The rise of wide-field, high-cadence photometric surveys such as Zwicky Transient Facility (ZTF)~\cite{2019PASP..131a8002B, 2019PASP..131f8003B} 
and the upcoming Vera C.\ Rubin Observatory Legacy Survey of Space and Time (LSST)~\cite{2009arXiv0912.0201L} has sparked efforts to develop scalable, photometry-only strategies for identifying unresolved lensed supernovae. Several methods have been proposed to meet this challenge. \cite{Goldstein:2017bny} used a template-fitting approach incorporating lensing priors. \cite{Bag:2020pbg} introduced a Bayesian model in which lensed supernovae are treated as the superposition of two time delayed and magnified copies of a smooth intrinsic light curve. \cite{Denissenya:2021cpz}  extended this with freeform deviations from the Hsiao supernova Ia spectral template~\cite{2007ApJ...663.1187H} for time delay modeling. This was successfully extended to shorter time delays and lower $S/N$ with deep learning in \cite{Denissenya:2022eds}. However, these studies primarily relied on simulated data, and their performance on real photometric data remains untested.

Recent observational and theoretical studies have examined the cosmological potential of strongly lensed supernovae. \cite{Linder:2004hx} discussed how time delays from strong lensing can complement other probes such as supernovae and the CMB to constrain dark energy, while \cite{Pierel:2019pnr} analyzed the potential of lensed events for $H_0$ and other cosmological parameter constraints. Building on these developments, efforts have also been made to improve the identification and exploitation of lensed supernovae in large surveys. For example, \cite{Liao:2019qoc} proposed model-independent cosmological inference using Gaussian processes applied to lensing time delays, and other studies have demonstrated the use of multiply imaged supernovae for direct $H_0$ measurements~\cite{Kelly:2023mgv}. Recently, \cite{Townsend:2024oyz} searched the ZTF archive for candidate lensed supernovae Ia using variable source selection, thereby motivating flexible detection approaches that do not rely on specific supernova templates or detailed lens models. An alternative strategy is to search for supernovae in previously identified strong-lensing systems identified in lens surveys, rather than performing blind searches across the transient population~\cite{Khalouei:2025jsw}.

In this work, we build upon the framework introduced by \cite{Bag:2020pbg}, employing the same combination of Chebyshev polynomial basis and Crossing Statistics~\cite{Shafieloo:2012jb} to model the intrinsic supernova light curve flexibly, but applying it for the first time to real observational data. Our method assumes that the light curve of an unresolved, strongly lensed supernova can be represented as the superposition of two time delayed, magnified copies of a single smooth light curve. We use Hamiltonian Monte Carlo (HMC) to infer the joint posterior distribution of the lensing parameters, magnification ratio $\mu$ and time delay $\Delta t$, while marginalizing over light curve shape parameters. Importantly, our approach is template-free and does not assume any specific supernova explosion model. The flexibility of the Chebyshev expansion enables its application to a broad range of transient types.

A key strength of our method is that it does not require spatially resolved multiple images, morphological classification, or spectroscopic redshift information. This independence from high-resolution imaging or spectroscopic follow-up makes the approach particularly suitable for wide-field, ground-based transient surveys, where the angular separation between multiple lensed images can be smaller than the typical seeing limit and therefore unresolved. In such cases, traditional lens detection techniques that rely on image multiplicity or morphological distortions are ineffective, and spectroscopic redshifts are often unavailable due to the prohibitive follow-up cost for large numbers of faint transients expected from upcoming surveys.

Since strongly lensed supernovae that produce multiple images with detectable time delays are intrinsically rare, stringent control of the purity, or false positive rate, is essential. Although wide-field surveys will detect millions of transient sources, only a subset of these are classified as supernovae, and an even smaller fraction satisfy the quality and sampling requirements necessary for lensing analysis. It would not be useful to flag thousands of candidates for follow-up without high confidence in their lensing nature. Instead, we aim to identify potentially strongly lensed supernovae whose inferred light curve structure indicates a high probability of multiple imaging and measurable time delay, thereby maximizing the scientific and cosmological return of follow-up observations.

To develop practical selection criteria, we apply our method to a sample of 524 spectroscopically confirmed Type Ia supernovae from the ZTF DR2 release~\cite{Rigault:2024kzb}. Because the probability of strong lensing for individual supernovae is small, this sample can be treated as effectively unlensed for the purpose of validating the false positive behavior of the pipeline.

Applying our method to this dataset allows us to evaluate how often unlensed light curves are identified as lensed candidates, thereby providing an empirical estimate of the contamination level expected in photometry-only searches. At the same time, the analysis helps us identify recurring failure modes, such as implausible flux ratios and marginal $\Delta t$ values, and refine selection thresholds for posterior convergence, time delay estimates, and $\chi^2$ improvements across the $g$ and $r$ bands. This work provides important insight into the behavior and limitations of the photometry-only pipeline under realistic observational conditions.

Our results show that this model-independent lensing inference pipeline is robust and computationally tractable for large transient datasets, making it well-suited for next-generation surveys such as LSST. By directly applying a methodology previously explored in simulation studies to real observational data~\cite{Bag:2020pbg}, our work provides an essential bridge between theoretical investigations and practical survey applications, paving the way toward the systematic discovery and cosmological use of lensed supernovae.

In \cref{sec:Data}, we describe the ZTF photometric dataset used in this work, including the sample of spectroscopically confirmed Type Ia supernovae employed for validation. In \cref{sec:Methodology}, we present our model-independent, photometry-based framework for identifying strongly lensed supernovae, outlining the light curve modeling procedure and the Bayesian HMC inference. In \cref{sec:results}, we apply this method to the ZTF validation sample and quantify both the statistical and physical false positive rates. In \cref{sec:discussion}, we provide a unified discussion of the method’s performance, outline its limitations and possible improvements, and examine implications for future wide-field surveys such as LSST. Finally, in \cref{sec:summary}, we summarize our findings and highlight directions for future work.

\section{Data}
\label{sec:Data}

In this section, we describe the ZTF survey and the subset of supernova light curves used to validate our photometric lensing framework.

\subsection{Zwicky Transient Facility Survey}
\label{sec:ZTF}

The Zwicky Transient Facility (ZTF) is a wide-field optical time-domain survey carried out with the 48-inch Schmidt Telescope at Palomar Observatory. Its large $47~\mathrm{deg}^2$ field of view and rapid cadence enable high-cadence monitoring of large areas of the northern sky~\cite{2019PASP..131a8002B,2019PASP..131f8003B}. ZTF uses a dedicated 600-megapixel camera to obtain multi-band imaging in $g$, $r$, and $i$, reaching median depths of $g \approx 20.8$~mag and $r \approx 20.6$~mag. Photometry is generated through an image-differencing pipeline and provided as forced PSF photometry on the difference images~\cite{2019PASP..131a8003M}, yielding uniform, well-calibrated light curves suitable for time-domain analysis.

\subsection{ZTF SN Ia DR2 Catalog}
\label{sec:ZTF2}

This work uses the ZTF SN Ia DR2 release of~\cite{Rigault:2024kzb}, which provides light curves for spectroscopically confirmed Type Ia supernovae. DR2 represents one of the largest homogeneous low-redshift SN Ia samples available, with consistent photometric processing and spectroscopic classifications. The survey spans multiple years of observations, enabling the construction of a large, homogeneous sample of well-observed SN Ia light curves that is well-suited for testing strong-lensing search methods based solely on unresolved light curves.

\subsection{Sample Selection}
\label{sec:Samples}

We use the ZTF SN Ia DR2 sample of~\cite{Rigault:2024kzb}, which provides 3,592 spectroscopically confirmed Type Ia supernovae with publicly available light curves\footnote{\url{https://ztfcosmo.in2p3.fr/}}. Although ZTF light curves include observations in the $g$, $r$, and $i$ bands, we restrict our analysis to the $g$ and $r$ bands due to their substantially higher signal-to-noise ratios.

To construct a validation dataset with reliable temporal sampling, we apply a simple quality cut based on light curve coverage around peak brightness. For each band ($g$ and $r$), we estimate the epoch of maximum flux and require sufficient sampling both before and after this peak. Specifically, we require at least 10 data points within the interval $[-20,0]$ days relative to the peak and at least 20 data points within $[0,+40]$ days, considering only observations with flux greater than 15\% of the peak flux in that band. Both conditions must be satisfied independently in the $g$ and $r$ bands.

These criteria ensure adequate coverage of the rising and declining phases of the light curve, which is necessary for robust two-component HMC fits. This selection is used solely to define a high-quality validation subset for evaluating the false positive behavior of our photometric lensing framework.

The peak epoch in each band is determined using an iterative smoothing algorithm \cite{Shafieloo:2005nd, Shafieloo:2007cs, Shafieloo:2009hi, Aghamousa:2014uya} applied to the light curves. The smoothing procedure and the fitting window used in the subsequent modeling are described in detail in \cref{sec:fitting}.

Applying these requirements yields 524 supernovae Ia with sufficient sampling for reliable two-component modeling.

\section{Methodology}
\label{sec:Methodology}

This work aims to identify strongly lensed supernovae from unresolved light curves using a model-independent method. In principle, gravitational lensing can produce multiple images of a supernova, which may lead to temporally separated features in the observed flux. In rare cases where the time delays are sufficiently long, lensing can be identified by the appearance of distinct, well-separated peaks in the multi-image light curve or by the discovery of new point sources that are temporally and spatially clustered. Although more than two images are possible in general, galaxy-scale lenses are often dominated by two-image configurations, which are the focus of this work.

However, in many realistic lensing configurations, the expected time delays are of order days to weeks, and the resulting light curves may appear as a single blended signal rather than as clearly distinguishable peaks. Detecting lensing signatures or measuring time delays in such unresolved cases is therefore intrinsically challenging, particularly in the absence of prior knowledge about whether an event is lensed.

The difficulty of this task depends on the supernova type. Type Ia supernovae (SNe Ia), owing to their relatively homogeneous light curve shapes, are well described by existing templates and therefore provide a controlled setting for time-domain analyses. For other supernova classes with greater intrinsic diversity, light curve modeling and time delay inference remain more challenging. Moreover, searches for lensed supernovae must contend with the large number of supernova light curves already discovered in existing surveys, as well as the rapidly increasing data volumes anticipated from ongoing and future wide-field surveys such as LSST.

Relying solely on distinct peaks in light curves is not a viable strategy, because such peaks occur only for unusually large time delays and are therefore exceedingly rare. This would lead to very low completeness for realistic lensing configurations. Photometry-only approaches that can identify unresolved lensed events are therefore important.

In this paper, we explore the feasibility of identifying strongly lensed supernovae from unresolved light curves by applying the model-independent framework proposed by~\cite{Bag:2020pbg} to real observational data. Specifically, we validate the method using a sample of spectroscopically confirmed, unlensed Type Ia supernovae, enabling a direct assessment of the false positive behavior of the approach. The following sections describe our sample selection, light curve modeling, fitting procedure, and prior choices used in the analysis. Applications to real lensed supernova candidates will be presented in a subsequent paper.

\subsection{Modeling the Lightcurve}
\label{sec:Modeling}

Strong gravitational lensing produces multiple images of a background source. In many cases, however, these images cannot be spatially resolved and instead appear as a single blended signal. As a result, the observed light curve corresponds to the superposition of the individual lensed images. Since all images originate from the same source, they share the same intrinsic light curve shape but differ in magnification factors $a_i$ and time delays $t_i$ due to their distinct light paths through the lensing potential. We assume that all images experience the same dust extinction along their paths.

The resulting blended light curve in the $j$-th filter, summed over $N_i$ images, can be expressed as

\begin{equation}
F_j(t) = \sum_{i = 1}^{N_i}a_if_j(t-t_i),
\label{eq:light1}
\end{equation}
where $f_j$ represents the intrinsic light curve in the $j$-th filter. Following the formulation by~\cite{Bag:2020pbg}, the general parameters of each lensed image can be characterized by the relative magnification $\mu_i \equiv a_i / a_1$ and the relative time delay $\Delta t_i \equiv t_i-t_1$. Throughout this work, we neglect the effects of microlensing, which can induce wavelength-dependent magnification variations. Under this assumption, the relative magnification parameters $\mu_i$ are assumed identical across all filters. With this parameterization and setting $t_1 = 0$ without loss of generality, \cref{eq:light1} becomes:

\begin{equation}
F_j(t) = \sum_{i = 1}^{N_i}\mu_i\mathcal{F}_j(t-\Delta t_i),
\label{eq:light2}
\end{equation}
where $\mathcal{F}_j(t)$ denotes the normalized intrinsic light curve in the $j$-th filter. The model also naturally accommodates the unlensed supernova case, corresponding to $N_i = 1$, or equivalently $\Delta t_i = 0$ and $\mu_i = 0$ for all but one component. Although strong lenses can produce more than two images, components with small magnification or nearly identical delays contribute negligibly to unresolved blended photometry, and most lensing information is effectively encoded in the two dominant images. For this reason, we restrict our inference to a two-image hypothesis in practice. Previous work has also demonstrated successful recovery of lensing signals in simulations of four-image systems~\cite{Denissenya:2021cpz,Denissenya:2022eds}.

This form, \cref{eq:light2}, models the observed light curve as a sum of scaled and time-shifted versions of a common intrinsic light curve shape $\mathcal{F}(t)$, modulated by the relative magnification $\mu$ and the time delay $\Delta t$. The primary challenge lies in choosing an appropriate functional form for $\mathcal{F}(t)$. Unlike quasars, whose variability is stochastic and complex, supernova light curves evolve on day timescales and follow a characteristic, smooth temporal evolution. They start faint, rise rapidly to a peak due to the explosion, and then gradually decline over time. This evolution should be accurately captured in the model for effective inference of lensing parameters. 

Therefore, we model $\mathcal{F}(t)$ as a log-normal function multiplied by a basis expansion involving the first four Chebyshev polynomials:

\begin{align}
\mathcal{F}_j(t) &= N_j \frac{1}{t} \exp\left[ -\frac{(\ln{t} - b_j)^2}{2\sigma_j^2} \right] \nonumber \\
&\quad \times \Big[ 1 
+ C_{1,j} t_{s,j} 
+ C_{2,j}(2t_{s,j}^2 - 1) \label{eq:chev}
\\
&\qquad 
+ C_{3,j}(4t_{s,j}^3 - 3t_{s,j}) 
+ C_{4,j}(8t_{s,j}^4 - 8t_{s,j}^2 + 1) \Big]. \nonumber 
\end{align}
where $b_j$ and $\sigma_j$ respectively control the location and width of the light curve peak in logarithmic time, and $N_j$ sets the overall normalization in filter $j$. The Chebyshev coefficients $C_{k,j}$ provide deviations from the baseline log-normal shape, and are not intended to define the light curve independently. In practice, unlensed supernova light curves are typically well described by the baseline log-normal profile with only small low-order corrections. However, the lensing-induced blended structure generally requires higher-order distortions to be reproduced accurately, motivating the inclusion of Chebyshev terms up to fourth order, as adopted in previous work~\cite{Bag:2020pbg}. 

Here, $t$ is measured in days, with $t=0$ corresponding to the start of the fitting range. To avoid numerical instabilities near the start of the fitting interval, we replace $t$ by $t + t_{\mathrm{floor}}$ in the log-normal envelope, with $t_{\mathrm{floor}} = 0.5\,\mathrm{days}$. This small offset stabilizes the evaluation of $\ln t$ at early times and has a negligible effect on the inferred light curve shape.

The scaled time variable is defined as $t_{s,j} \equiv t / t_{j,\mathrm{max}} - 1$, where $t_{j,\mathrm{max}}$ is the endpoint of the fitting interval in filter $j$. Although Chebyshev polynomials are formally orthogonal over the interval $[-1,1]$, we restrict the domain to $t_{s,j} \in [-1,0]$ in order to suppress oscillatory behavior in the fitted light curves. In practice, using the full $[-1,1]$ interval tends to produce wiggly fits that do not resemble realistic supernova light curves. Restricting the domain to $[-1,0]$ yields significantly smoother fits and therefore acts as an effective regularization against overfitting.

Therefore,~\cref{eq:chev} models the intrinsic light curve as a log-normal profile capturing the characteristic rise and decay of supernovae, with a Chebyshev polynomial expansion allowing for controlled, finite deviations from the canonical shape. Chebyshev polynomials, widely used in Crossing Statistics, provide a compact and efficient basis for modeling small departures from a fiducial light curve without introducing excessive flexibility~\cite{Shafieloo:2010xm, Shafieloo:2012jb, Shafieloo:2012yh, Hazra:2014hma}.

Finally, adopting the two-image model discussed above, the blended light curve becomes

\begin{equation}
F_j(t) = \mathcal{F}_j(t) + \mu\mathcal{F}_j(t - \Delta t),
\label{eq:twoimage}
\end{equation}
where $\mu$ and $\Delta t$ represent the relative magnification and time delay between the two images. Both components share the same intrinsic light curve shape as given in \cref{eq:chev}.

In a naive implementation of \cref{eq:twoimage}, the delayed component contributes no flux for $t < \Delta t$ and a finite contribution for $t > \Delta t$. This introduces a sharp turn-on at $t=\Delta t$, making the model non-differentiable with respect to $\Delta t$.
To avoid this issue, we replace the sharp turn-on with a smooth gating function. Specifically, the contribution of the delayed image is multiplied by a logistic gating factor that smoothly transitions from zero to unity across $t\approx\Delta t$, and the delayed light curve is evaluated at a softened time coordinate that smoothly approaches $(t-\Delta t)$ at late times. This treatment preserves the physical interpretation of a delayed second image while avoiding non-differentiable behavior that would otherwise hinder efficient sampling.

\setlength{\tabcolsep}{7pt}
\renewcommand{\arraystretch}{1.2}
\begin{table*}[ht]
\centering
\caption{\justifying
Prior distributions adopted for all model parameters. Subscripts $g$ and $r$ denote the parameters for the two photometric bands. Flux-related parameters are defined in dimensionless units after the normalization described in~\cref{sec:fitting}. All priors are weakly informative and chosen to ensure numerical stability while preserving broad physical flexibility.
}
\label{tab:priors}

\begin{tabular}{lll}
\hline
Parameter & Prior Distribution & Description \\
\hline
$N_{\mathrm{param},g}$, $N_{\mathrm{param},r}$ &
$\mathrm{LogNormal}(\ln N_{0j},\,0.4)$ &
Band–dependent flux normalization \\

$b_g$, $b_r$ &
$\mathcal{N}(3.5,\,0.5)$ &
Log-time location of the light curve peak \\

$\sigma_g$, $\sigma_r$ &
$\mathrm{LogNormal}(\ln 0.5,\,0.4)$ &
Light curve width \\

$C_{k,g}, C_{k,r}$ &
$\mathcal{N}(0,\,5)$ &
Chebyshev coefficients ($k=1,\dots,4$) \\

$\mu$ &
$\mathrm{LogNormal}(0,\,0.6)$ &
Relative magnification \\

$\Delta t$ &
$\mathcal{N}(10,\,50)$ on $[5,50]$ &
Time delay (days) \\
\hline
\end{tabular}
\end{table*}

\subsection{Fitting the Lightcurve}
\label{sec:fitting}

Before fitting, it is critical to define an appropriate time range of the data over which the light curve model in~\cref{eq:twoimage} is fitted. This choice can significantly impact the inferred lensing parameters, as different temporal selections of the observed data can lead to substantial variations in the fit. We emphasize that this procedure pertains only to the selection of observed data points and does not modify the light curve model itself.

To robustly determine the fitting range, we adopt an iterative smoothing algorithm following~\cite{Shafieloo:2005nd, Shafieloo:2007cs, Shafieloo:2009hi, Aghamousa:2014uya}. The smoothed light curve is used only to identify the overall trend and locate the flux peak, thereby defining a stable temporal window for data selection. Importantly, the smoothed light curve is not used in the likelihood evaluation or parameter inference; all fits are performed directly on the original photometric data.

At each iteration step $n$, the smoothed light curve $F^{s}_{n}(t)$ is updated from the previous step $n-1$ via
\begin{multline}
F^{s}_{n}(t) = F^{s}_{n-1}(t) + N(t) \sum_{i} \dfrac{F_{\textrm{obs}}(t_i) - F^{s}_{n-1}(t_i)}{\sigma_{\textrm{obs}}^2(t_i)} \\
\times \exp \left[-\dfrac{(t_i - t)^2}{2\Delta^2} \right],
\label{eq:iter_lightcurve}
\end{multline}
where the normalization factor $N(t)$ is defined as

\begin{equation}
N(t)^{-1} = \sum_{i} \exp \left[-\dfrac{(t_i - t)^2}{2\Delta^2} \right] \dfrac{1}{\sigma_{\textrm{obs}}^2(t_i)}.
\label{eq:norm_lightcurve}
\end{equation}

Here, $F_{\textrm{obs}}(t_i)$ and $\sigma_{\textrm{obs}}(t_i)$ represent the observed flux and its associated uncertainty at time $t_i$, and $F^{s}_{0}(t)$ denotes the initial guess. The kernel width $\Delta$ determines the degree of smoothing, i.e., how strongly neighboring data points influence the reconstructed light curve at a given time $t$.  

We apply this smoothing independently to the $g$ and $r$ bands to estimate the peak flux time in each band. Based on the identified peak, we first define a band-specific reference fitting window extending from $-40$ to $+50$ days relative to the peak (observer frame), restricted to epochs where the smoothed flux exceeds $15\%$ of the peak value. This reference window is chosen to capture the characteristic rise and decline phases of the supernova light curve. To ensure robustness against uncertainties in the peak estimation and to allow sufficient fitting flexibility near the boundaries, we further extend each band-specific window by $\pm10$ days.

For the joint multi-band analysis, we define a global fitting window whose start and end times are given by the earliest band-specific start time and the latest band-specific end time, respectively. All observations within this global window are included in the fit. The two bands are not required to have measurements at identical epochs. Each band contributes data only within its available temporal coverage, while the lensing parameters are jointly constrained using both bands.

Once the final fitting range is established, we infer the lensing parameters $\mu$ and $\Delta t$ by fitting the model in~\cref{eq:twoimage} to the observed light curves within a Bayesian inference framework. Assuming Gaussian photometric uncertainties, the likelihood is given by

\begin{equation}
\mathcal{L} \propto
\exp\left(-\frac{1}{2}\chi^2\right),
\end{equation}
where

\begin{equation}
\chi^2 =
\sum_i
\frac{\left[F_{\mathrm{obs}}(t_i)-F_{\mathrm{model}}(t_i)\right]^2}
{\sigma_{\mathrm{obs}}^2(t_i)}.
\end{equation}

We have also tested a Cauchy likelihood and find that it yields similar results, but with slightly less stable convergence. Therefore, we adopt the Gaussian form here. Posterior sampling is performed using Hamiltonian Monte Carlo (HMC), as implemented in the \texttt{pystan} package~\cite{JSSv076i01}.

In this setup, the lensing parameters $\mu$ and $\Delta t$ are inferred jointly using both $g$ and $r$ bands, while the intrinsic light curve shape parameters are fitted independently for each band. Given the high dimensionality and correlations among the shape parameters, HMC is well-suited for sampling the resulting posterior distribution.

We evaluate two competing models for each supernova light curve:

\begin{itemize}
\item a no-lensing model, in which each band is described by a single intrinsic light curve with no magnified or time delayed component; and
\item a lensing model, in which $\mu$ and $\Delta t$ are free parameters shared across bands, while the shape parameters remain band-specific.
\end{itemize}

For each posterior draw, we evaluate the $\chi^2$ statistic from the data--model residuals. For the lensing model, this is done jointly for the two-band fit, yielding band-specific contributions $\chi^2_g$ and $\chi^2_r$ as well as the total statistic $\chi^2_{\rm tot}=\chi^2_g+\chi^2_r$. For the no-lensing model, the $g$ and $r$ band light curves are fitted independently, and the total $\chi^2$ is constructed as the sum of the band-specific values.

For visualization, we define representative posterior draws: a ``best'' draw corresponding to the minimum $\chi^2$ value and a ``median'' draw whose $\chi^2$ is closest to the posterior median.

Although the no-lensing configuration formally corresponds to special cases within the lensing model (e.g., $\mu \rightarrow 0$ or $\Delta t \rightarrow 0$), we treat the two models as distinct hypotheses in practice. This approach avoids relying on poorly sampled regions of the lensing parameter space near these boundaries and provides a more stable basis for our false positive tests.

A source is classified as a potential lensing candidate only if the lensing model yields a lower $\chi^2$ than the no-lensing model in both the $g$ and $r$ bands separately, thereby reducing false positives caused by improvements confined to a single band.

Before passing the data to the sampler, we apply the same iterative smoothing procedure used to define the fitting range (see \cref{eq:iter_lightcurve}) to estimate the overall trend of each light curve and to identify its peak flux. The observed fluxes and their associated uncertainties in the $g$ and $r$ bands are then rescaled by a common normalization factor defined as the larger of the smoothed peak fluxes in the two bands. This shared normalization preserves relative band scaling while yielding dimensionless light curves with typical amplitudes of order unity. All flux and uncertainty values entering the Bayesian inference are consistently normalized in this way, improving numerical stability during sampling by keeping the flux amplitudes and their uncertainties close to unity, thereby avoiding large parameter scales during HMC exploration and simplifying the interpretation of the amplitude parameters. The normalization is applied only for model fitting and posterior sampling. For visualization and presentation, the model evaluated on representative posterior draws, together with the data, is rescaled to the original flux units using the same normalization factors, so that all figures are shown in physically meaningful observational units.

\subsection{Prior Selection}
\label{sec:prior}

We adopt weakly informative priors for all model parameters, motivated by typical properties of Type~Ia supernova light curves and the need for stable HMC sampling. The full set of priors is summarized in \cref{tab:priors}.

Briefly, amplitude and width parameters are assigned lognormal priors to enforce positivity while allowing order-unity variations. The peak-location parameters are modeled with Gaussian priors centered on characteristic Type~Ia time-scales, while the Chebyshev coefficients are given broad Gaussian priors to allow flexible deviations without introducing unphysical behavior.

The relative magnification $\mu$ is also assigned a lognormal prior, ensuring positivity while remaining only weakly informative. For the time delay, we adopt a truncated normal prior over $\Delta t \in [5,50]$ days. The lower bound excludes the near-degenerate regime $\Delta t \rightarrow 0$, where $\mu$ and $\Delta t$ become poorly identifiable and lead to unstable inference.

\setlength{\tabcolsep}{7pt}
\renewcommand{\arraystretch}{1.2}
\begin{table*}[ht]
\centering
\caption{\justifying
Summary of the final candidate satisfying the adopted primary selection criterion $\Delta t \ge 12$ days. The full set of objects passing the more permissive threshold $\Delta t \ge 10$ days is provided in Appendix Table~\ref{tab:lens_candidates_all}. Reported values of $\mu$ and $\Delta t$ are posterior medians with 68\% credible intervals derived from the posterior samples of the representative seed. The quantities $\Delta\chi^2_g$ and $\Delta\chi^2_r$ denote the median-based differences between the lensed and unlensed fits in the $g$ and $r$ bands, respectively. $\Delta \mathrm{DIC}_{\rm tot}$ denotes the total median-deviance DIC difference. All $\Delta$ quantities are defined as $\Delta X \equiv X_{\rm lensed} - X_{\rm unlensed}$, so that negative values favor the lensed model. The final column gives the fraction of divergent transitions in the Stan HMC chains for the representative seed used to compute the reported summary statistics.}
\label{tab:lens_candidates}
\begin{tabular}{cccccccccc}
\hline
\multicolumn{10}{c}{\textbf{Final candidate} ($\Delta t \ge 12$ days)} \\
\hline
No. & Name & $\mu$ & $\Delta t$ [days] & $\hat{R}_\mu$ & $\hat{R}_{\Delta t}$ & $\Delta\chi^2_g$ & $\Delta\chi^2_r$ & $\Delta \mathrm{DIC}_{\rm tot}$ & Div. \\
\hline
$\#1$ & ZTF20abnwldu & $0.46^{+0.03}_{-0.04}$ & $12.57^{+0.28}_{-0.31}$ & 1.000 & 0.999 & $-44.79$ & $-19.58$ & $-65.17$ & 0.000 \\
\hline
\end{tabular}
\end{table*}

\section{Results and Analysis}
\label{sec:results}

In this section, we first describe the selection procedure and convergence criteria used to identify lensed supernova candidates. We then assess the false positive behavior of the pipeline using a large control sample of spectroscopically confirmed Type~Ia supernovae. This sample is used to validate the overall procedure and empirically quantify the false positive rate, rather than to tune hyperparameters or decision thresholds.

\subsection{Selection Criteria and Convergence Filtering}
\label{sec:selection}

To ensure robust inference and to minimize false positives in the identification of strongly lensed supernovae, we adopt a set of selection criteria motivated by previous work~\cite{Bag:2020pbg,Denissenya:2021cpz}. In particular, \cite{Bag:2020pbg} demonstrated that while a threshold of $\Delta t \ge 10$ days yields low false positive rates, a small level of contamination ($\sim$8\%) can remain depending on the noise properties. Crucially, increasing the threshold to $\Delta t \ge 12$ days completely eliminates false positives in their simulations, with none of the unlensed supernovae being misclassified as unresolved lensed events.

Motivated by this result, we adopt $\Delta t \ge 12$ days as our primary acceptance criterion to construct a high-purity sample. The $\Delta t \ge 10$ day threshold is retained only as a reference to illustrate how the candidate sample changes under a more permissive cut.

We apply our joint two-band Hamiltonian Monte Carlo (HMC) inference framework to a control sample of 524 spectroscopically confirmed Type~Ia supernovae from the ZTF DR2 sample~\cite{Rigault:2024kzb}. For each object, we fit the $g$ and $r$ band light curves simultaneously using four independent HMC chains of 2,000 iterations each, discarding the first 1,000 iterations per chain as warm-up. Posterior summaries and convergence diagnostics are computed using only the post–warm-up samples.

We first require that the lensing parameters, the relative magnification $\mu$ and the time delay $\Delta t$, satisfy the Gelman-Rubin convergence criterion~\cite{10.1214/ss/1177011136} with $\hat{R}<1.05$, consistent with the standard adopted in~\cite{Bag:2020pbg}. We further exclude fits with significant HMC pathologies by requiring that the fraction of divergent transitions remain below $2\%$. For consistency with the reported summary statistics, this diagnostic is evaluated using the representative seed for each object, rather than aggregated over all runs.

To reduce the impact of occasional sampling pathologies in HMC, each supernova is analysed using five independent random seeds. In complex posterior landscapes, HMC trajectories can sometimes explore different local regions of parameter space depending on the initialization, which may lead to inconsistent fits across runs. Therefore, we require that the convergence and selection criteria be satisfied in at least three of the five runs. This majority-vote requirement ensures that candidate detections are stable against stochastic sampling effects.

For reference, the numbers of objects achieving convergence in at least $1/5$, $2/5$, $3/5$, $4/5$, and $5/5$ seeds are 500, 473, 445, 419, and 359 out of 524 objects, respectively.

Finally, among the converged fits, we classify an object as a lensed supernova candidate only if it satisfies all of the following criteria, which are predefined and applied uniformly to the entire control sample:

\begin{enumerate}

\item A measured time delay satisfying
\[
\Delta t \ge 12~\mathrm{days},
\]
a regime in which false positive rates are expected to be negligible based on previous simulation results~\cite{Bag:2020pbg}.

\item Improvement of the lensed model fit in both bands, quantified by negative differences in the data-only chi-square statistic
\[
\Delta\chi^2_g < 0, \qquad
\Delta\chi^2_r < 0,
\]
where $\Delta\chi^2$ is computed from the likelihood term only and does not include prior contributions.

\item Preference for the lensed model under the deviance information criterion (DIC),
\[
\Delta\mathrm{DIC}_{\rm tot} < 0,
\]
where the DIC provides a Bayesian measure of model quality that balances goodness-of-fit and effective model complexity.

\item Posterior consistency of the inferred lensing parameters, requiring
\[
P(\Delta t \ge 12\,\mathrm{days}) \ge 0.95
\]
and
\[
P(1/3 \le \mu \le 3) \ge 0.95 ,
\]

\end{enumerate}
which corresponds to relative magnifications of order unity typical of unresolved galaxy-scale lensing configurations, and helps suppress solutions associated with poorly constrained or degenerate fits that would otherwise increase the number of statistical false positives.

We emphasize that the $\Delta t \ge 12$ day criterion is chosen to ensure a high-purity sample with negligible false positives. Objects with $10 \le \Delta t < 12$ days are therefore treated as intermediate or marginal cases and are not included in the final candidate sample. As will be shown in the validation analysis, the majority of statistically selected objects are concentrated in this intermediate regime, indicating that false positives predominantly arise near the selection boundary.

Not all objects yield well-converged HMC fits, even under the selection framework described above. In our control sample, 445 out of 524 supernovae achieve convergence in at least three of the five runs. The remaining cases are typically characterized by limited temporal coverage, low signal-to-noise ratios, or sparse sampling near peak light, resulting in weakly constrained posteriors.

In such cases, the data provide little information on the lensing parameters, and the posterior distributions remain broad and poorly constrained. Therefore, these systems fail the convergence and robustness criteria and are excluded from the candidate classification step. 

We note that this behavior is consistent with scenarios in which the observed light curves are adequately described by a single-component model, such that introducing a second delayed component does not lead to a stable or well-constrained fit. However, we do not explicitly classify these systems as unlensed, as the lack of convergence may also reflect limitations in data quality rather than the absence of lensing.

\subsection{Validation on Unlensed Supernovae}
\label{sec:validation}

The converged subset of the control sample provides a direct empirical estimate of the false positive rate of our selection procedure. Because these objects are spectroscopically confirmed Type~Ia supernovae and are not expected to be strongly lensed, any object that satisfies the selection criteria can be interpreted as a false positive.

Applying the full selection criteria described in~\cref{sec:selection}, including the adopted threshold $\Delta t \ge 12~\mathrm{days}$, to the 445 converged fits leaves only a single object satisfying all conditions. This corresponds to a false positive rate of $1/445 \approx 0.22\%$. This result is consistent with the findings of~\cite{Bag:2020pbg}, who showed that such a threshold eliminates false positives in simulations.

For comparison, if the more permissive threshold $\Delta t \ge 10~\mathrm{days}$ is adopted, the same procedure yields fourteen objects, corresponding to a statistical false positive rate of $14/445 \approx 3.15\%$. This represents the contamination level associated with the 10-day threshold.

This result is fully consistent with the findings of~\cite{Bag:2020pbg}, who showed that while a $\Delta t \ge 10$ day threshold yields low but non-zero false positive rates, increasing the threshold to $\Delta t \ge 12$ days removes false positives entirely. Our analysis demonstrates the same trend in real data. The candidate sample is substantially reduced when moving from 10 to 12 days, with only one object remaining as a robust candidate.

Importantly, this validation sample is used exclusively to evaluate the false positive behavior of the fixed selection criteria, with no threshold tuning or optimization performed on this dataset, ensuring that the reported false positive rates represent an unbiased evaluation of the selection procedure.

\begin{figure*}

\centering

\includegraphics[width = 0.8\linewidth]{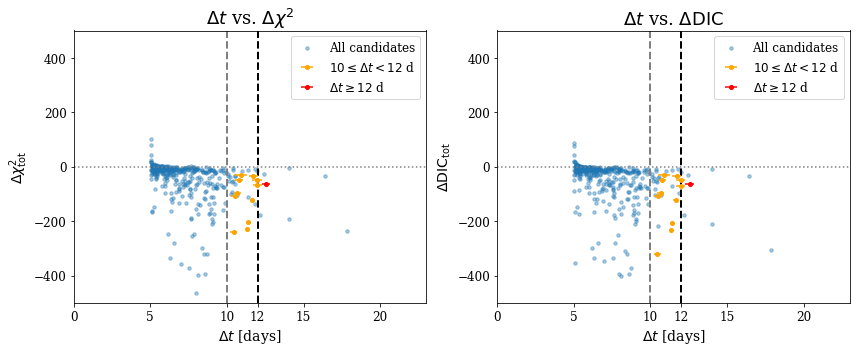}
\caption{\label{fig:fig_dt_vs_chi2_dic} \justifying
Selection diagnostics for a validation sample of spectroscopically confirmed Type Ia supernovae. \textit{Left:} $\Delta t$ versus $\Delta\chi^2_{\rm tot}$. \textit{Right:} $\Delta t$ versus $\Delta{\rm DIC}_{\rm tot}$. Blue points show all candidates with at least three converged seed fits ($N=445$). Orange points indicate candidates in the intermediate range $10 \leq \Delta t < 12$ days ($N=13$). The single red point marks the final candidate satisfying the $\Delta t \geq 12$ day criterion ($N=1$). The vertical dashed gray and black lines mark $\Delta t=10$ and $12$ days, respectively, and the horizontal dotted gray lines indicate $\Delta\chi^2_{\rm tot}=0$ and $\Delta{\rm DIC}_{\rm tot}=0$.
}

\end{figure*}

\begin{figure}

\centering

\includegraphics[width = 0.9\linewidth]{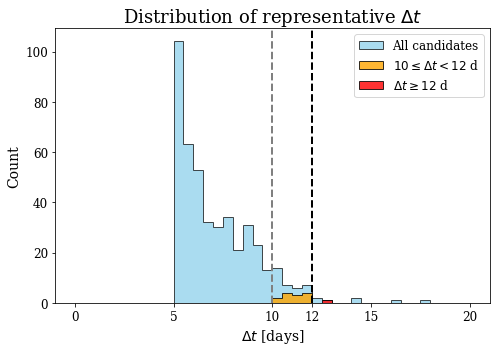}
\caption{\label{fig:dist_dt} \justifying
Distribution of the inferred time delay $\Delta t$ for the validation sample. The blue histogram shows all candidates with at least three converged seed fits ($N=445$). The orange histogram denotes objects in the intermediate range $10 \leq \Delta t < 12$ days ($N=13$), while the red histogram marks the single final candidate with $\Delta t \geq 12$ days ($N=1$). The dashed gray and black vertical lines indicate $\Delta t=10$ and $12$ days, respectively.
}
  
\end{figure}

\begin{figure}

\centering

\includegraphics[width = 0.9\linewidth]{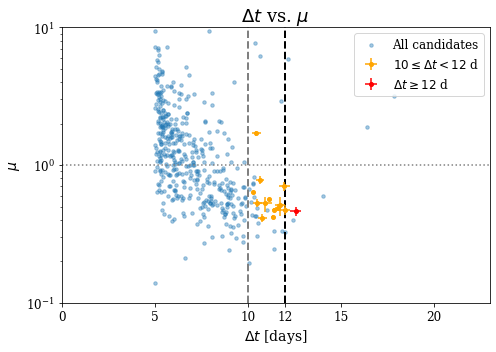}
\caption{\label{fig:fig_dt_vs_mu} \justifying
Time delay $\Delta t$ versus flux ratio $\mu$ for the validation sample. Blue points show all candidates with at least three converged seed fits ($N=445$). Orange points indicate objects in the intermediate range $10 \leq \Delta t < 12$ days ($N=13$), while the red point marks the final candidate with $\Delta t \geq 12$ days ($N=1$). Error bars represent the central 68\% posterior intervals. The vertical dashed gray and black lines mark $\Delta t=10$ and $12$ days, respectively, and the horizontal dotted gray line indicates $\mu=1$.
}
\end{figure}

\begin{figure*}
\centering

\subfloat[$\#1$ ZTF20abnwldu, unlensed]
{\includegraphics[width=0.48\linewidth]{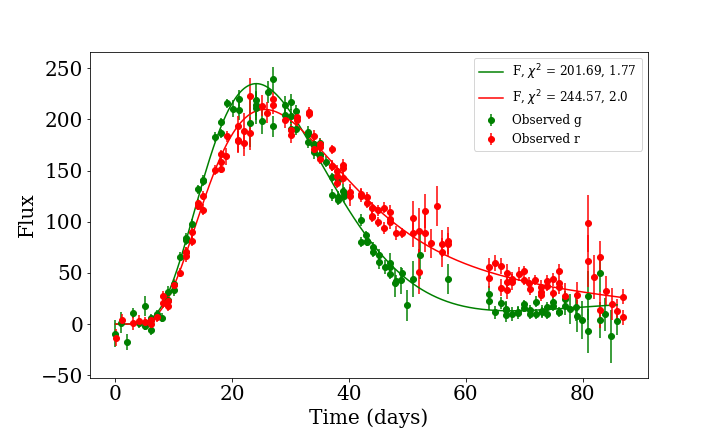}}
    \label{fig:1a}
\hspace{1em}
\subfloat[$\#1$ ZTF20abnwldu, lensed]
{\includegraphics[width=0.48\linewidth]{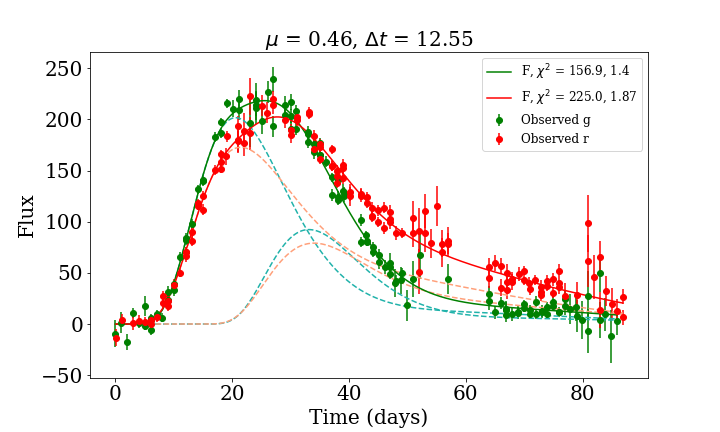}}
    \label{fig:1b}

\caption{\label{fig:lightcurve1} \justifying
Representative unlensed (\textit{left}) and lensed (\textit{right}) light curve fits for the final candidate ($\#1$ ZTF20abnwldu) identified from the validation sample of spectroscopically confirmed Type~Ia supernovae. This object satisfies the criteria, including $\Delta t \ge 12$ days, and represents the most robust candidate in the sample. The plotted curves correspond to a representative posterior realization, selected to be closest to the median $\chi^2$ value among the posterior samples. The values of $\mu$ and $\Delta t$ displayed above the lensed panel correspond to the parameters of this plotted realization. In the lensed panel, dashed curves denote the two inferred components, $f_1$ and $f_2$, and the solid curve shows their summed contribution. The legend reports the total $\chi^2$ and reduced $\chi^2$ values for each fit.
}

\end{figure*}

\subsection{Lensing Candidate Identification}
\label{sec:candidates}

Building on the validation results, we examine the properties of the events that satisfy the selection criteria and assess their interpretation in the context of false positive contamination.

As established in \cref{sec:validation}, only a single event satisfies the full selection criteria when adopting the primary threshold $\Delta t \ge 12$ days. We therefore identify this object as the final candidate. In addition, thirteen objects fall in the intermediate range $10 \le \Delta t < 12$ days, which we interpret as marginal cases arising from the more permissive threshold.

The $\Delta t$--$\Delta\chi^2_{\rm tot}$ and $\Delta t$--$\Delta{\rm DIC}_{\rm tot}$ projections in \cref{fig:fig_dt_vs_chi2_dic} illustrate how the selection criteria isolate these subsets within the broader population. The single red point corresponds to the final candidate satisfying $\Delta t \ge 12$ days, while the orange points denote objects with $10 \le \Delta t < 12$ days. 
Although the selection is applied separately in the $g$ and $r$ bands, the figure shows the combined statistic $\Delta\chi^2_{\rm tot}$ for clarity. Taken together, these results indicate that the selection acts as a joint constraint in parameter space, rather than being driven by any single diagnostic.

In addition, \cref{fig:dist_dt} shows the distribution of posterior-median $\Delta t$ values for all 445 converged events, as well as for the intermediate and final candidates. Both distributions decline toward larger $\Delta t$, so that only a small fraction of events exceed $\Delta t = 10$ days, and even fewer extend beyond $\Delta t = 12$ days. 
This demonstrates that the time delay cut primarily acts as a boundary in parameter space, and that tightening the threshold significantly reduces contamination.

The properties of the final candidate and intermediate objects are further illustrated in \cref{fig:fig_dt_vs_mu}, which shows the relationship between the inferred time delay $\Delta t$ and flux ratio $\mu$. By construction, these objects satisfy the posterior constraint on the flux ratio, while spanning a range of $\mu$ values within the allowed interval. This constraint excludes solutions in which one component dominates the combined light curve and the second contributes only a weak perturbation. Together with the time delay requirement $\Delta t \ge 12$ days, this places the final candidate in a regime where the two-component structure remains observationally distinguishable, even in unresolved data. No strong correlation between $\Delta t$ and $\mu$ is apparent, indicating that the selection is driven primarily by temporal structure in the light curves rather than by flux-ratio constraints.

Given the validation sample of 445 well-converged supernovae, the identification of a single candidate when applying the baseline requirement $\Delta t \ge 12$ days corresponds to a false positive fraction of 
\[
f_{\mathrm{FP},{\rm baseline}} = \frac{1}{445} \approx 0.22\%\ .
\] 
If one employs the laxer criterion of 
$\Delta t \ge 10$ days (which was cautioned against by \cite{Bag:2020pbg}), 
one obtains fourteen objects, or a false positive fraction of $14/445=3.15\%$. 
This reflects the fact that most false positives are concentrated near the selection boundary, and demonstrates that a stricter time delay cut significantly improves sample purity. 

In practice, statistically selected candidates require additional astrophysical vetting beyond the scope of this statistical framework. Such follow-up would assess whether the inferred parameters are consistent with physically plausible strong-lensing configurations, and is essential for distinguishing genuinely lensed systems from the intrinsically ambiguous cases identified here.

\subsection{Light Curve Behavior of the Final Candidate}
\label{sec:lc_examples}

To illustrate the nature of the selected events, \cref{fig:lightcurve1} shows the unlensed and lensed light curve fits for the final candidate identified in this work. The plotted curves correspond to representative posterior realizations selected to be close to the median $\chi^2$ values, and therefore may not coincide exactly with the posterior summary statistics reported in \cref{tab:lens_candidates}.

The final candidate ($\#1$ ZTF20abnwldu) is the only object that passes all the criteria, including $\Delta t \ge 12$ days (see \cref{sec:candidates}). In this regime, the lensed model introduces a temporal offset between two components, resulting in a more structured light curve and an improved statistical fit relative to the unlensed model.

However, the inferred components are not cleanly separated in time, and the resulting light curve does not exhibit an unambiguous two-peak structure. Instead, the second component remains blended with the primary one, modifying the overall shape rather than producing a clearly distinct feature. This behavior reflects the intrinsic limitation of unresolved photometric data, where even for $\Delta t \sim 12$ days the two components can remain significantly overlapped.

The intermediate objects with $10 \le \Delta t < 12$ days (shown in Appendix~\cref{fig:lightcurve2,fig:lightcurve3,fig:lightcurve4,fig:lightcurve5}) show qualitatively similar behavior, with the inferred components remaining strongly blended. 
In such cases, the lensed model accommodates small deviations from a single-component evolution, presumably 
fitting noise in the data to obtain a better $\chi^2$.

\section{Discussion and Future Outlook}
\label{sec:discussion} 

The Vera C. Rubin Observatory’s Legacy Survey of Space and Time (LSST) will discover millions of supernovae with multi-band, high-cadence photometry, far exceeding the capacity for spectroscopic or high-resolution follow-up. In this setting, scalable photometric pre-selection methods become relevant, since many lensed supernovae will appear as unresolved, blended light curves. Our analysis shows that a template-free, photometry-only framework can serve as an effective first-stage filter.

Because under realistic survey conditions, photometry alone cannot conclusively confirm lensing, this method is best interpreted as the first stage of a hierarchical selection pipeline. The role of the photometric selection is to identify a well-defined subset of candidates that exhibit lens-like features, while acknowledging that a fraction of these will remain ambiguous. These candidates can then be prioritized for targeted follow-up, including spectroscopy, host-galaxy redshift measurements, or high-resolution imaging with facilities such as HST, JWST, or ground-based adaptive optics.

In addition, our inference framework imposes strict convergence and consistency requirements that go beyond standard template-based approaches such as SALT2. As a result, some supernovae that would be accepted under conventional light curve fitting are rejected in our analysis due to unstable or poorly constrained posterior behavior. This highlights that the method functions not only as a model comparison framework but also as a robustness filter that helps exclude unreliable fits.

Finally, the model-independent nature of this approach enables application across a wide range of transient classes, including those lacking reliable templates. This flexibility will be particularly valuable for LSST, where the diversity of discovered events will far exceed current samples. Together, these considerations demonstrate that photometric lensing searches can serve as an efficient, scalable component of a broader transient-lensing discovery pipeline for next-generation surveys.

\section{Summary}
\label{sec:summary} 

We have presented and validated a model-independent, photometry-only framework for identifying lensed candidates in supernova light curves. Using a two-component representation of the observed flux and Hamiltonian Monte Carlo inference for the relative scaling and time delay, the method avoids reliance on supernova templates or lensing mass models and can be applied broadly across diverse transient populations.

To quantify the behavior of this approach under realistic observational conditions, we applied it to a validation sample of spectroscopically confirmed unlensed Type~Ia supernovae. Of the 445 objects with well-converged posterior inferences, only a single object satisfies the full selection criteria when adopting the primary threshold of $\Delta t \ge 12$ days, corresponding to a false positive fraction of $1/445 \approx 0.22\%$ (a threshold of $\Delta t \ge 10$ days would yield fourteen objects, or 3.15\% false positive fraction). 

The framework developed here provides a practical, scalable first-stage filter for large-scale time-domain surveys. By identifying events that deviate from single-component light curve behavior, the method can significantly reduce the volume of data requiring follow-up. When combined with targeted spectroscopic or high-resolution imaging observations, this approach provides a viable pathway to discover and confirm strongly lensed supernovae in next-generation surveys such as LSST. In addition, we are poised to apply the finalized pipeline to larger, unlabeled ZTF samples, enabling a systematic search for real lensed supernova candidates using the validated selection framework.

\section*{Acknowledgments}

This work was supported by the high performance computing cluster Seondeok at the Korea Astronomy and Space Science Institute (KASI).
This work was supported by the U.S. Department of Energy (DOE), Office of Science, Office of High Energy Physics, under Contract No. DE–AC02–05CH11231. This work made use of data accessed through the DESI–ZTF cross-collaboration agreement.

\section*{Data Availability}

This work uses publicly available light curves from the ZTF SN Ia DR2 sample, available at \url{https://ztfcosmo.in2p3.fr/}.

\appendix
\section{Intermediate Objects and Additional Light Curve Fits}
\label{app:lightcurves}

\setlength{\tabcolsep}{7pt}
\renewcommand{\arraystretch}{1.2}
\begin{table*}[ht]
\centering
\caption{\justifying
Summary of the objects passing the statistical selection with the more permissive threshold $\Delta t \ge 10$ days. The single final candidate satisfying $\Delta t \ge 12$ days is listed separately from the thirteen intermediate objects with $10 \le \Delta t < 12$ days. Reported values of $\mu$ and $\Delta t$ are posterior medians with 68\% credible intervals derived from the posterior samples of the representative seed. The quantities $\Delta\chi^2_g$ and $\Delta\chi^2_r$ denote the median-based differences between the lensed and unlensed fits in the $g$ and $r$ bands, respectively, and $\Delta \mathrm{DIC}_{\rm tot}$ denotes the total median-deviance DIC difference. All $\Delta$ quantities are defined as $\Delta X \equiv X_{\rm lensed} - X_{\rm unlensed}$, so that negative values favor the lensed model. The final column gives the fraction of divergent transitions in the Stan HMC chains.
}
\label{tab:lens_candidates_all}
\begin{tabular}{cccccccccc}
\hline
No. & Name & $\mu$ & $\Delta t$ [days] & $\hat{R}_\mu$ & $\hat{R}_{\Delta t}$ & $\Delta\chi^2_g$ & $\Delta\chi^2_r$ & $\Delta \mathrm{DIC}_{\rm tot}$ & Div. \\
\hline
\multicolumn{10}{c}{\textbf{Final candidate} ($\Delta t \ge 12$ days)} \\
\hline
$\#1$ & ZTF20abnwldu & $0.46^{+0.03}_{-0.04}$ & $12.57^{+0.28}_{-0.31}$ & 1.000 & 0.999 & -44.79 & -19.58 & -65.17 & 0.000 \\
\hline
\multicolumn{10}{c}{\textbf{Intermediate objects} ($10 \le \Delta t < 12$ days)} \\
\hline
$\#2$ & ZTF19abdqapq & $0.53^{+0.06}_{-0.07}$ & $10.91^{+0.40}_{-0.43}$ & 1.000 & 1.000 & -25.43 & -4.76 & -31.59 & 0.050 \\
$\#3$ & ZTF20aayxkqy & $0.47^{+0.04}_{-0.04}$ & $11.98^{+0.31}_{-0.33}$ & 1.000 & 1.000 & -5.19 & -42.93 & -49.38 & 0.000 \\
$\#4$ & ZTF19aavidqf & $0.42^{+0.01}_{-0.01}$ & $11.33^{+0.10}_{-0.11}$ & 0.999 & 1.000 & -144.59 & -84.95 & -231.50 & 0.000 \\
$\#5$ & ZTF20aaiaaoi & $0.70^{+0.05}_{-0.05}$ & $11.97^{+0.29}_{-0.28}$ & 1.000 & 1.000 & -59.46 & -9.48 & -70.52 & 0.075 \\
$\#6$ & ZTF18abobkii & $0.51^{+0.08}_{-0.08}$ & $11.71^{+0.30}_{-0.27}$ & 1.000 & 1.000 & -15.45 & -19.53 & -35.68 & 0.075 \\
$\#7$ & ZTF20acberfr & $0.78^{+0.05}_{-0.05}$ & $10.67^{+0.21}_{-0.23}$ & 1.000 & 1.000 & -33.68 & -64.51 & -98.57 & 0.000 \\
$\#8$ & ZTF19aaujzqh & $0.57^{+0.01}_{-0.01}$ & $11.15^{+0.06}_{-0.06}$ & 1.002 & 1.004 & -458.85 & -1458.71 & -2025.25 & 0.000 \\
$\#9$ & ZTF20acuzlib & $0.53^{+0.03}_{-0.03}$ & $10.51^{+0.24}_{-0.26}$ & 1.001 & 0.999 & -12.94 & -94.84 & -108.80 & 0.000 \\
$\#10$ & ZTF20aaghnxd & $0.42^{+0.03}_{-0.03}$ & $10.78^{+0.26}_{-0.25}$ & 1.002 & 1.002 & -2.93 & -45.23 & -49.72 & 0.000 \\
$\#11$ & ZTF18abryewg & $0.47^{+0.02}_{-0.02}$ & $11.39^{+0.09}_{-0.09}$ & 1.001 & 0.999 & -49.25 & -153.63 & -205.46 & 0.000 \\
$\#12$ & ZTF19abirbpd & $1.70^{+0.07}_{-0.07}$ & $10.46^{+0.23}_{-0.24}$ & 1.001 & 1.001 & -176.93 & -64.74 & -321.03 & 0.075 \\
$\#13$ & ZTF19aanmdsr & $0.49^{+0.03}_{-0.03}$ & $11.65^{+0.15}_{-0.16}$ & 1.000 & 1.000 & -68.41 & -55.27 & -121.07 & 0.000 \\
$\#14$ & ZTF19aamhhey & $0.64^{+0.01}_{-0.01}$ & $10.26^{+0.05}_{-0.05}$ & 1.000 & 1.000 & -580.76 & -1091.23 & -1695.19 & 0.050 \\
\hline
\end{tabular}
\end{table*}

In this appendix, we present light curve fits for the remaining intermediate objects from the validation sample of spectroscopically confirmed Type~Ia supernovae. A complete summary of these objects, together with the final candidate, is provided in \cref{tab:lens_candidates_all}. These objects satisfy the statistical selection under the more permissive $\Delta t \ge 10$ day threshold but do not meet the adopted baseline criterion $\Delta t \ge 12$ days, and are therefore not included in the final candidate sample.

\begin{figure*}
\centering

\subfloat[$\#2$ ZTF19abdqapq, unlensed]{\includegraphics[height=0.25\linewidth]{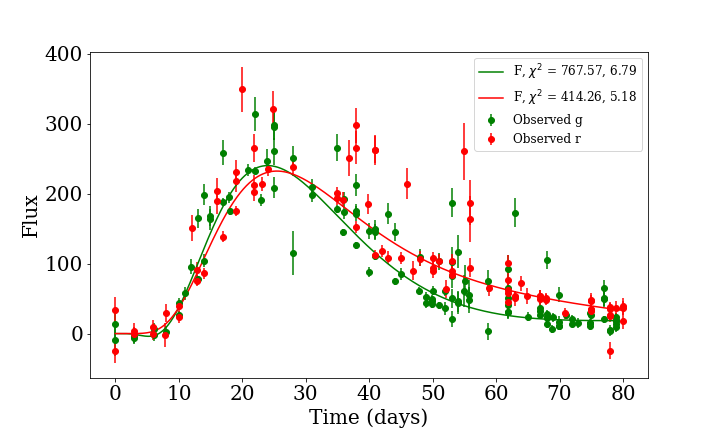}}
    \label{fig:2a}
\hspace{1em}
\subfloat[$\#2$ ZTF19abdqapq, lensed]{\includegraphics[height=0.25\linewidth]{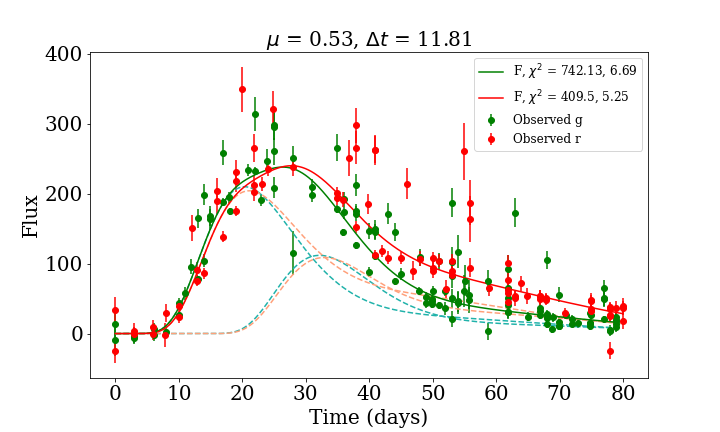}}
    \label{fig:2b}

\subfloat[$\#3$ ZTF20aayxkqy, unlensed]{\includegraphics[height=0.25\linewidth]{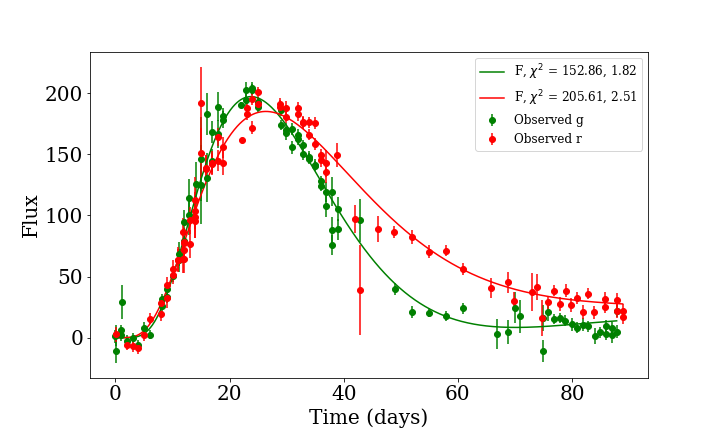}}
    \label{fig:3a}
\hspace{1em}
\subfloat[$\#3$ ZTF20aayxkqy, lensed]{\includegraphics[height=0.25\linewidth]{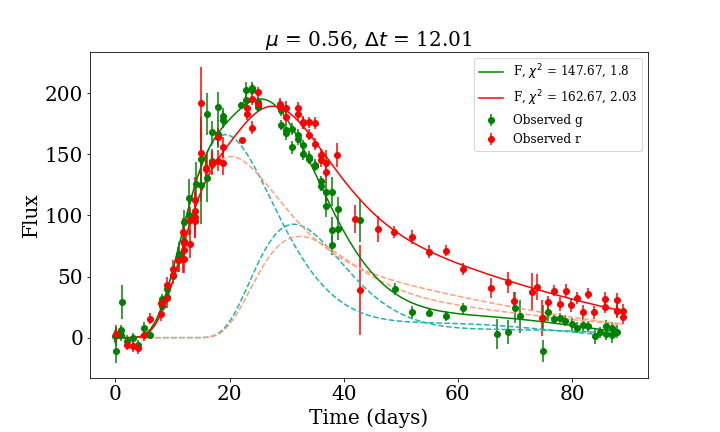}}
    \label{fig:3b}

\subfloat[$\#4$ ZTF19aavidqf, unlensed]{\includegraphics[height=0.25\linewidth]{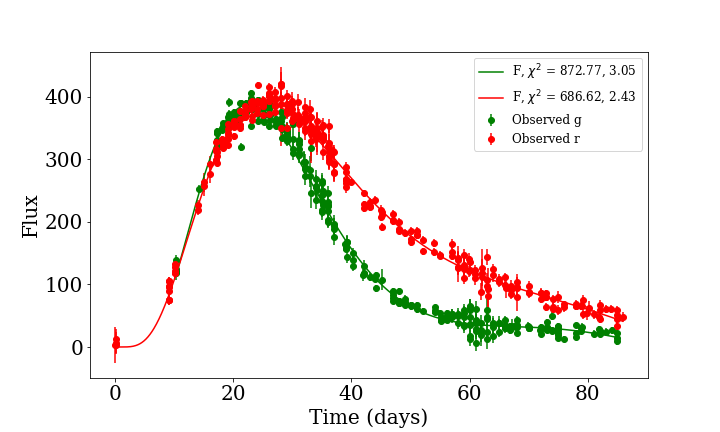}}
    \label{fig:4a}
\hspace{1em}
\subfloat[$\#4$ ZTF19aavidqf, lensed]{\includegraphics[height=0.25\linewidth]{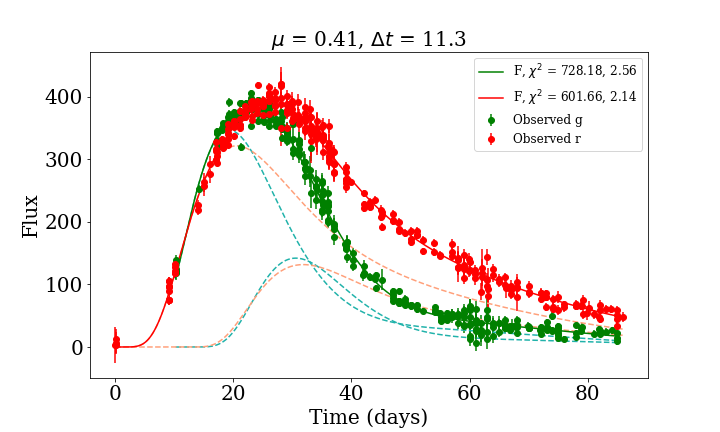}}
    \label{fig:4b}

\subfloat[$\#5$ ZTF20aaiaaoi, unlensed]{\includegraphics[height=0.25\linewidth]{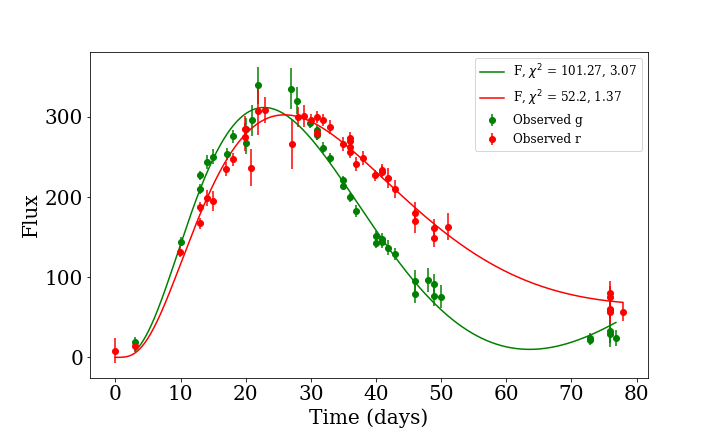}}
    \label{fig:5a}
\hspace{1em}
\subfloat[$\#5$ ZTF20aaiaaoi, lensed]{\includegraphics[height=0.25\linewidth]{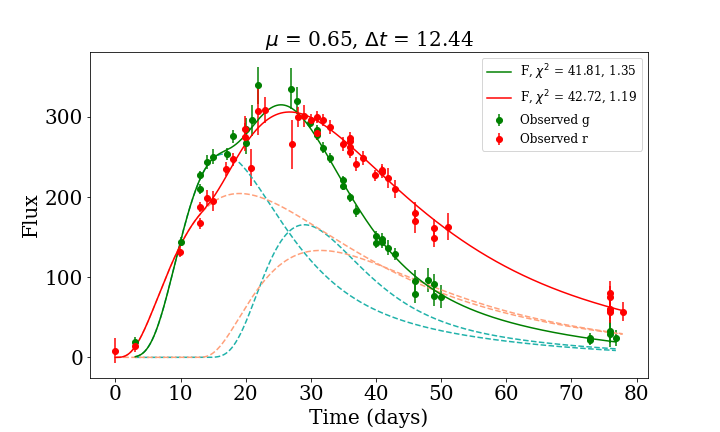}}
    \label{fig:5b}

\caption{\label{fig:lightcurve2} \justifying
Representative unlensed (\textit{left}) and lensed (\textit{right}) light curve fits for intermediate objects with $10 \le \Delta t < 12$ days from the validation sample of spectroscopically confirmed Type~Ia supernovae. These objects satisfy the statistical selection under the more permissive $\Delta t \ge 10$ day threshold but do not meet the adopted primary criterion $\Delta t \ge 12$ days, and are therefore treated as marginal cases. The plotted curves correspond to representative posterior realizations, selected to be closest to the median $\chi^2$ value among the posterior samples. The values of $\mu$ and $\Delta t$ displayed above each lensed panel correspond to the parameters of the plotted realizations. In the lensed panels, dashed curves denote the two inferred components, $f_1$ and $f_2$, and solid curves show their summed contribution. The legend reports the total $\chi^2$ and reduced $\chi^2$ values for each fit.
}

\end{figure*}

\begin{figure*}
\centering

\subfloat[$\#6$ ZTF18abobkii, unlensed]{\includegraphics[height=0.25\linewidth]{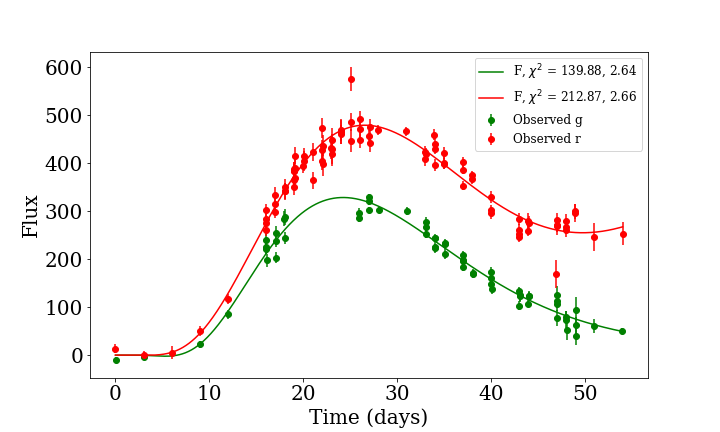}}
    \label{fig:6a}
\hspace{1em}
\subfloat[$\#6$ ZTF18abobkii, lensed]{\includegraphics[height=0.25\linewidth]{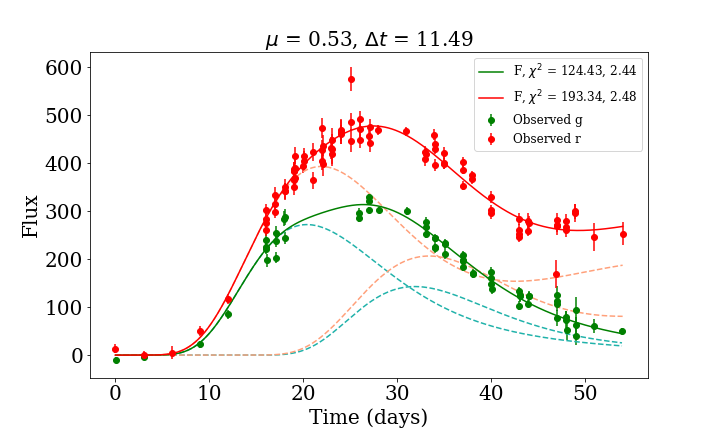}}
    \label{fig:6b}

\subfloat[$\#7$ ZTF20acberfr, unlensed]{\includegraphics[height=0.25\linewidth]{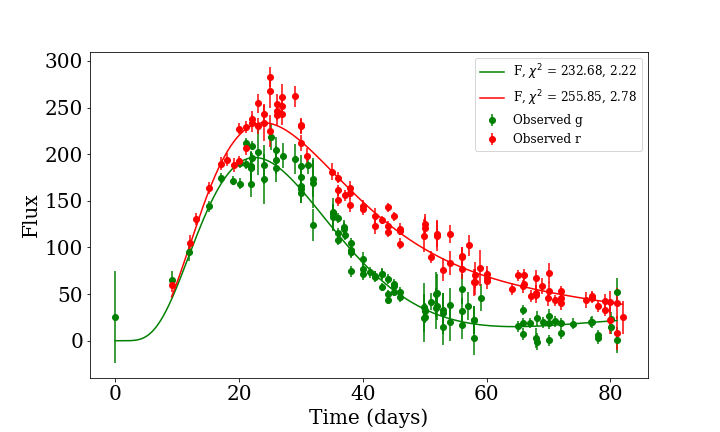}}
    \label{fig:7a}
\hspace{1em}
\subfloat[$\#7$ ZTF20acberfr, lensed]{\includegraphics[height=0.25\linewidth]{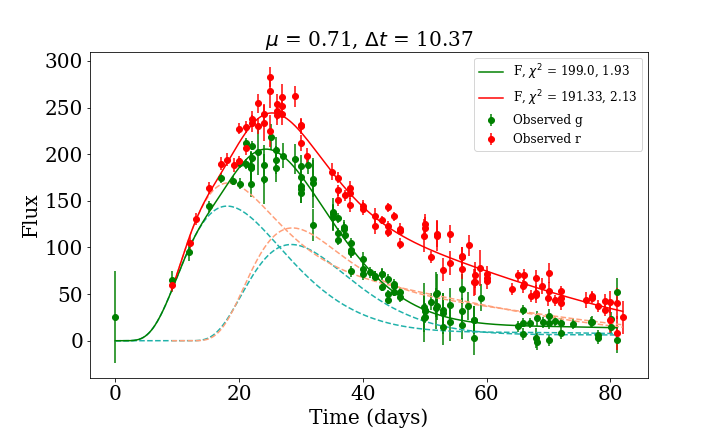}}
    \label{fig:7b}

\subfloat[$\#8$ ZTF19aaujzqh, unlensed]{\includegraphics[height=0.25\linewidth]{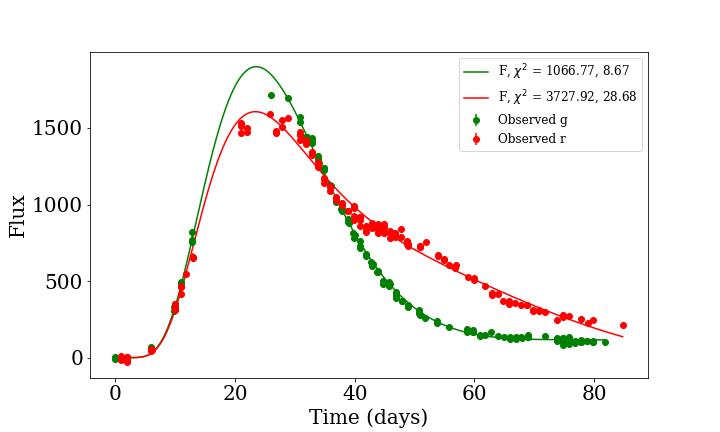}}
    \label{fig:8a}
\hspace{1em}
\subfloat[$\#8$ ZTF19aaujzqh, lensed]{\includegraphics[height=0.25\linewidth]{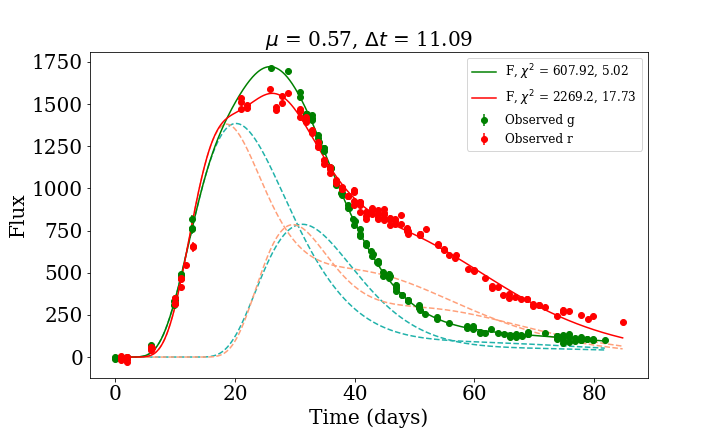}}
    \label{fig:8b}

\subfloat[$\#9$ ZTF20acuzlib, unlensed]{\includegraphics[height=0.25\linewidth]{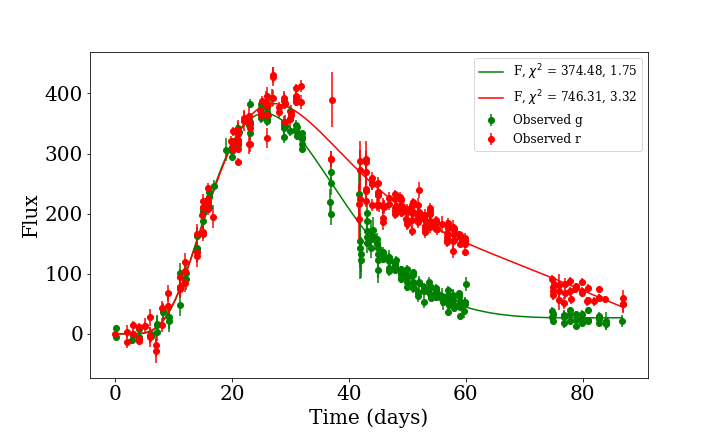}}
    \label{fig:9a}
\hspace{1em}
\subfloat[$\#9$ ZTF20acuzlib, lensed]{\includegraphics[height=0.25\linewidth]{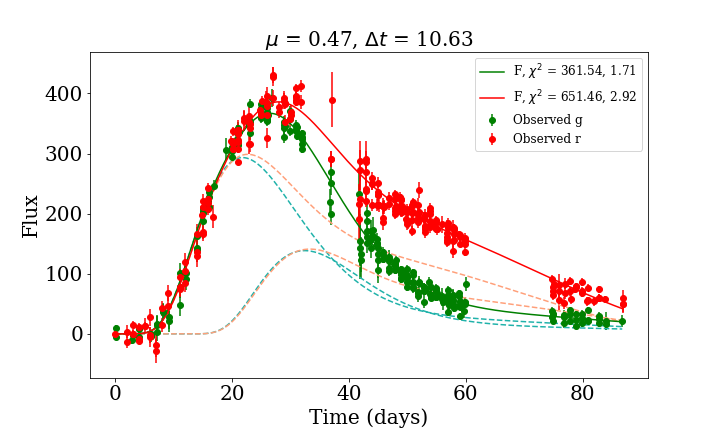}}
    \label{fig:9b}

\caption{\label{fig:lightcurve3} \justifying
(\textit{Continued}) Light curve fits for intermediate objects with $10 \le \Delta t < 12$ days.
}

\end{figure*}

\begin{figure*}
\centering

\subfloat[$\#10$ ZTF20aaghnxd, unlensed]{\includegraphics[height=0.25\linewidth]{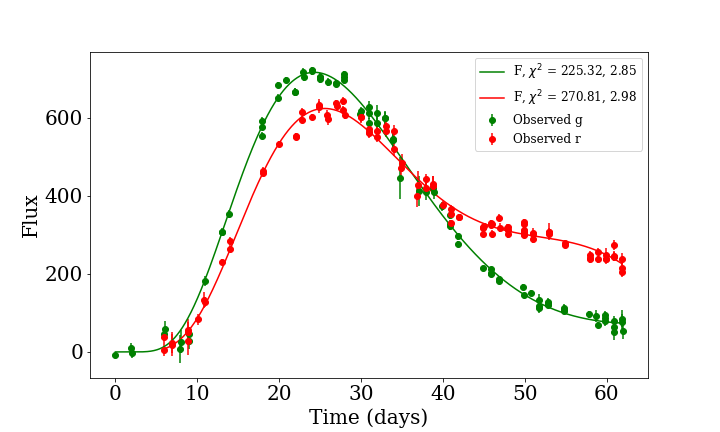}}
    \label{fig:10a}
\hspace{1em}
\subfloat[$\#10$ ZTF20aaghnxd, lensed]{\includegraphics[height=0.25\linewidth]{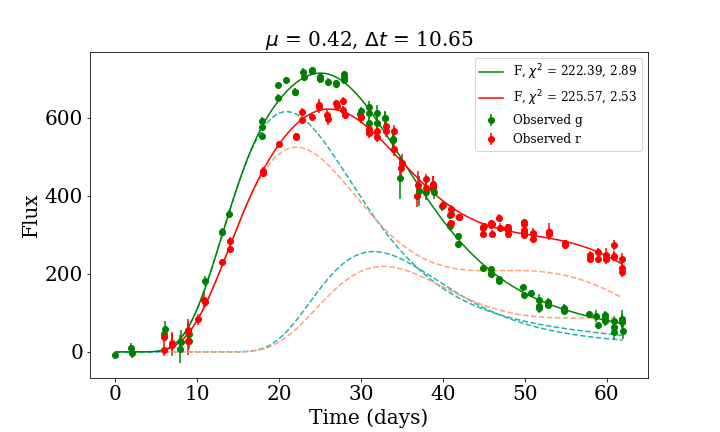}}
    \label{fig:10b}

\subfloat[$\#11$ ZTF18abryewg, unlensed]{\includegraphics[height=0.25\linewidth]{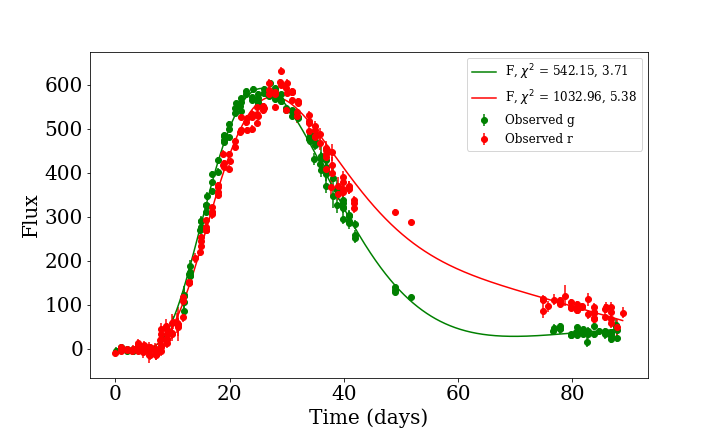}}
    \label{fig:11a}
\hspace{1em}
\subfloat[$\#11$ ZTF18abryewg, lensed]{\includegraphics[height=0.25\linewidth]{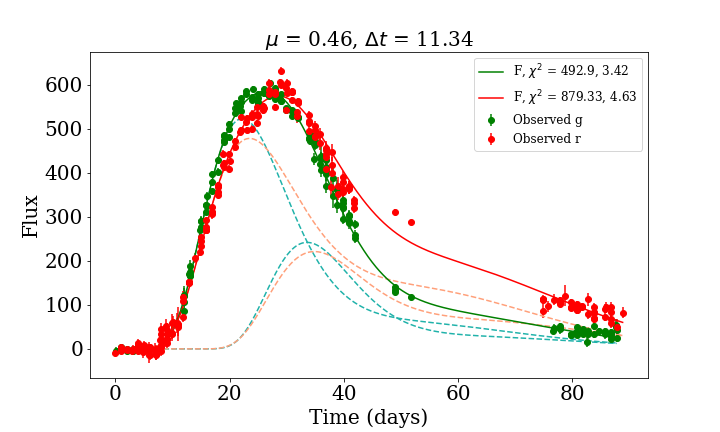}}
    \label{fig:11b}

\subfloat[$\#12$ ZTF19abirbpd, unlensed]{\includegraphics[height=0.25\linewidth]{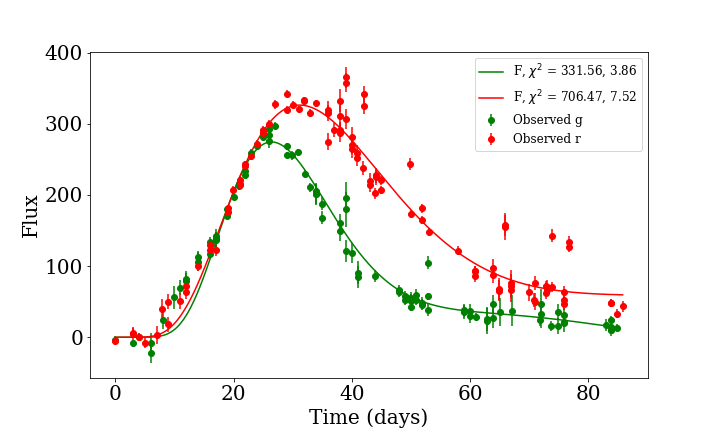}}
    \label{fig:12a}
\hspace{1em}
\subfloat[$\#12$ ZTF19abirbpd, lensed]{\includegraphics[height=0.25\linewidth]{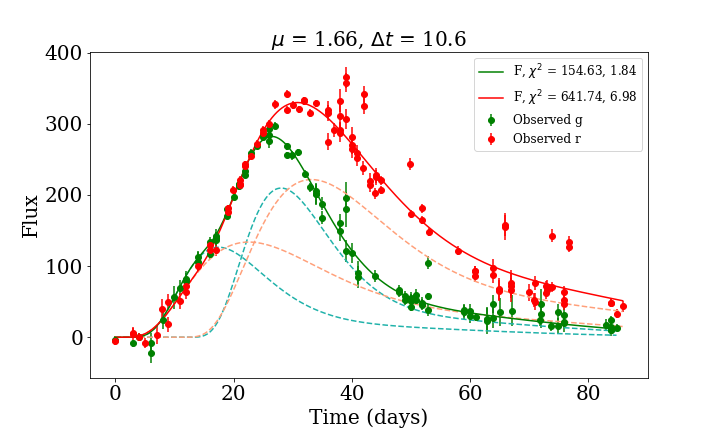}}
    \label{fig:12b}

\subfloat[$\#13$ ZTF19aanmdsr, unlensed]{\includegraphics[height=0.25\linewidth]{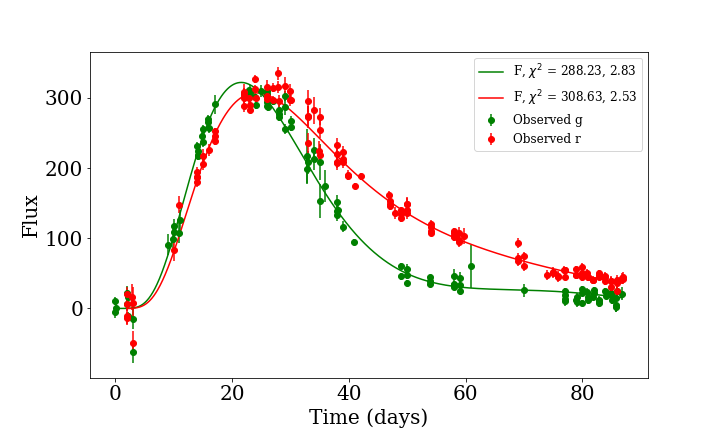}}
    \label{fig:13a}
\hspace{1em}
\subfloat[$\#13$ ZTF19aanmdsr, lensed]{\includegraphics[height=0.25\linewidth]{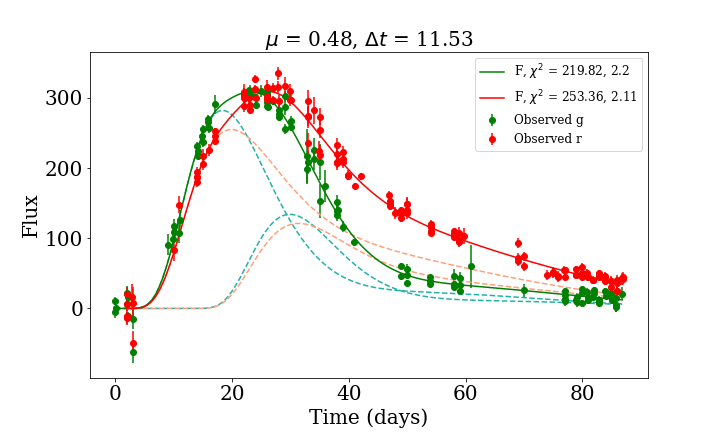}}
    \label{fig:13b}

\caption{\label{fig:lightcurve4} \justifying
(\textit{Continued}) Light curve fits for intermediate objects with $10 \le \Delta t < 12$ days.
}

\end{figure*}

\begin{figure*}
\centering

\subfloat[$\#14$ ZTF19aamhhey, unlensed]{\includegraphics[height=0.25\linewidth]{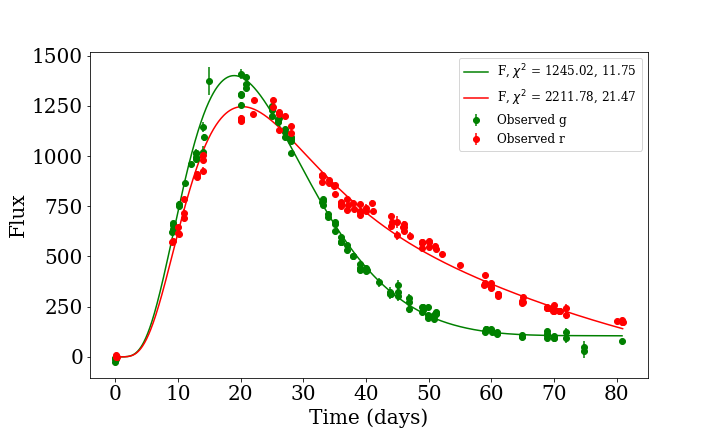}}
    \label{fig:14a}
\hspace{1em}
\subfloat[$\#14$ ZTF19aamhhey, lensed]{\includegraphics[height=0.25\linewidth]{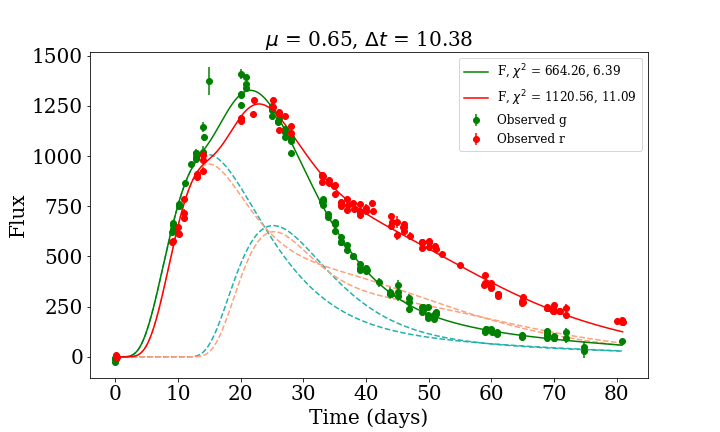}}
    \label{fig:14b}

\caption{\label{fig:lightcurve5} \justifying
(\textit{Continued}) Light curve fits for intermediate objects with $10 \le \Delta t < 12$ days.
}

\end{figure*}

The set of intermediate objects is shown in \cref{fig:lightcurve2,fig:lightcurve3,fig:lightcurve4,fig:lightcurve5}. 
Each object is shown with both the unlensed and lensed fits, allowing a direct comparison of how the two-component model improves the description of the data. As discussed in the main text, these objects are interpreted as statistical false positives arising from the application of the selection framework to an effectively unlensed population.

\bibliography{apssamp}

\end{document}